\documentstyle[aps,preprint]{revtex}
\tighten

\begin{document}

\title{ Effect of flavor mixing on the time delay
of massive supernova neutrinos}

\author{\it Sandhya Choubey\thanks{e-mail: sandhya@tnp.saha.ernet.in}
and Kamales Kar\thanks{e-mail: kamales@tnp.saha.ernet.in}\\
Saha Institute of Nuclear Physics,\\1/AF, Bidhannagar,
Calcutta 700064, INDIA.}
\maketitle



\def\nue{{\nu_e}}
\def\anue{{\bar\nu_e}}
\def\numu{{\nu_{\mu}}}
\def\anumu{{\bar\nu_{\mu}}}
\def\nutau{{\nu_{\tau}}}
\def\anutau{{\bar\nu_{\tau}}}
\newcommand{\dm}{\mbox{$\Delta{m}^{2}$~}}
\newcommand{\st}{\mbox{$\sin^{2}2\theta$~}}


\begin{abstract}
The neutrinos and antineutrinos of all the three flavors released from 
a galactic supernova will be detected in the water Cerenkov detectors. 
We show that even though the neutral current interaction is flavor blind, 
and hence neutrino flavor mixing cannot alter the total neutral current 
signal in the detector, it can have a non-trivial impact on
the delay of massive neutrinos and alters the neutral current event
rate as a function of time. We have suggested various variables of the 
neutral and charged current events that can be used to study this effect. 
In particular the ratio of charged to neutral current events can 
be used at early times while the 
ratio of the energy moments for the charged 
to the neutral current events can form useful diagnostic tools even 
at late times to study neutrino mass and mixing.  
\end{abstract}
\vskip 6mm



\vskip 8mm



The detection of the SN1987A neutrinos by the water 
Cerenkov detectors at Kamioka and IMB 
settled many important issues in  
the subject of type II supernova theory. The observation of neutrinos 
from any future 
galactic supernova event will answer the remaining 
questions regarding the understanding of the supernova mechanisms. A 
galactic supernova event will also bring in a lot of information 
on neutrino mass, which of late, has been an issue of much discussion. Among 
the various problems which demand non-zero neutrino mass are the 
atmospheric neutrinos anomaly \cite{atm,skatm}, the solar neutrino 
problem \cite{solar,sksolar} and the LSND experiment in Los Alamos \cite{lsnd}. 
While all the three above mentioned experiments give information on 
the mass squared differences, 
the supernova neutrinos can be used to place direct limits on the 
$\numu/\nutau$ masses and at the same time can also 
constrain the neutrino mixing parameters which will be 
useful in the understanding of the above three experiments. 

About $10^{58}$ neutrinos, in all three flavors 
carrying a few times $10^{53}$ ergs of energy are released in a type II 
supernova. These neutrinos for a galactic supernova events can be 
detected by the current water Cerenkov detectors, the Super-Kamiokande 
(SK) and the Sudbury Neutrino Observatory (SNO). The effect of 
neutrino mass can show up in the observed neutrino
signal in these detectors in two ways,
\begin{itemize}
\item by causing delay in the time of flight measurements
\item by modifying the neutrino spectra through neutrino flavor mixing
\end {itemize}

Massive neutrinos travel with speed less than the speed of light and 
for typical galactic supernova distances $\sim$ 10 kpc, 
even a small mass results in a measurable delay in the arrival time 
of the neutrino. Many different analyses have been performed before 
to give bounds on the neutrino mass (\cite{bv} and references 
therein). Neutrino oscillations on the other hand convert the more 
energetic $\numu/\nutau (\anumu/\anutau)$ into $\nue (\anue)$ thereby
hardening the resultant $\nue (\anue)$ energy spectra and 
hence enhancing their signal at the detector \cite{bkg2,qf,cmk}. 
In a previous work \cite{cmk} we studied quantitatively the effects
of neutrino
flavor oscillations on the supernova neutrino spectrum and the number of
charged current events at the detector using a realistic supernova model. 
In this work we study the neutral current signal
as a function of time 
in the water Cerenkov detectors, for a mass range of the neutrinos
where both the phenomenon of delay and flavor conversion are operative. 
That the time response of the event rate in the detector is 
modified if the neutrinos have mass alone and hence delay 
is a well known feature \cite{bv}. 
In this letter we stress the point that since 
neutrino flavor conversions change the energy spectra
of the neutrinos, and since 
the time delay of the massive neutrinos is energy dependent, 
the time dependence of the event rate at the detector
is altered appreciably in the presence of mixing. 
We suggest various variables which act as tools for  
measuring this change in the time response curve of the neutral 
current events and in differentiating the cases of (a) massless 
neutrinos (b) neutrinos with mass but no mixing and (c) neutrinos 
with mass as well as mixing. In particular 
we study the ratio of the charged current to neutral current ratio R(t), 
as a function of time in the SNO detector and show that the 
change in the value and the shape of R(t) due to flavor mixing 
cannot be emulated by 
uncertainties. We also study other variables like the normalized 
$n$-th energy moments of the neutral current events and the ratio of 
charged to the neutral current $n$-th moments as important diagnostic tools 
in filtering out the effects of neutrino mass and mixing.  

The differential number of neutrino events at the detector for a
given reaction process is
\begin{equation}
\frac{d^2 S}{dEdt}=\sum_i\frac{n}{4\pi D^2} 
N_{\nu_i}(t)f_{\nu_i}(E) \sigma(E)
\epsilon(E)
\label{sig}
\end{equation}
where $i$ runs over the neutrino species concerned, 
$N_{\nu_i}(t) = L_{\nu_i}(t)/\langle E_{\nu_i}(t)\rangle$, 
are the number of neutrinos produced at the source 
where $L_{\nu_i}(t)$ is the neutrino luminosity and
$\langle E_{\nu_i}(t) \rangle$ is the average energy, 
$\sigma (E)$ is the reaction cross-section for the neutrino with
the target particle, $D$ is the distance of the neutrino source
from the detector (taken as 10kpc),
$n$ is the number of detector particles for the reaction considered
and $f_{\nu_i} (E)$ is the energy spectrum for the neutrino species 
involved, 
while $\epsilon(E)$ is the detector efficiency as a function of the 
neutrino energy. 
By integrating out the energy from eq.(\ref{sig}) we get the time
dependence of the various reactions at the detector. To get the total
numbers both integrations over energy and time has to be done.

For the neutrino luminosities and average energies, though it is best to 
use a numerical supernova model, 
but for simplicity, we will here use a profile of the 
neutrino luminosities and temperatures which have general agreement 
with most supernova models. We take 
the total supernova energy radiated in neutrinos to be 3 $\times 10^{53}$ 
ergs. This luminosity, which is almost the same
for all the neutrino species, has a fast rise over a period of 0.1 sec
followed by a slow fall over several seconds in most 
supernova models. We use a luminosity that
has a rise of 0.1 sec using one side of the Gaussian with $\sigma$ = 0.03
and then an exponential decay with time constant $\tau$ = 3 sec for 
all the flavors \cite{bv}. 

The average energies associated with the $\nue, \anue ~\rm{and}~ \numu$ (the
$\numu, \anumu, \nutau ~\rm{and}~ \anutau$ have the same energy spectra)
are 11 MeV, 16 MeV and 25 MeV respectively in most numerical models.
We take these average energies and consider them to be
constant in time. We have also checked our calculations with
time dependent average energies and estimated its effect.
The neutrino spectrum is taken to be a pure Fermi-Dirac
distribution  characterized by the
neutrino temperature alone. 

We will here be concerned with two important water Cerenkov detectors, 
the SK and the SNO. 
The column 1 of Table 1 lists all the important reactions in SK and SNO.  
In column 2 of Table 1 we report the calculated number of 
expected events for the various reactions in SNO, 
when neutrinos are assumed to be massless. The corresponding 
values for SK can be obtained by scaling the number of events 
in $\rm H_2O$ to its fiducial volume of 32 kton. 
The detector efficiency is taken to be 1 and the 
energy threshold is taken to be 5 MeV for both SK and
SNO \cite{bv}. For the cross-section of the $(\nue-d), (\anue-d),
(\nu_i-d)$
and $(\anue-p)$ reactions we refer to \cite{burrows}.
The cross-section of the $(\nue(\anue)-e^-)$ and $(\nu_i-e^-)$
scattering has been taken from \cite{kolb} while the neutral current
$(\nu_i-^{16}O)$ scattering cross-section is taken from \cite{bv}.
For the
$^{16}O(\nue-e^-)^{16}F$ and $^{16}(\anue,e^+)^{16}N$ reactions we
refer to \cite{haxton} and use it's cross-sections
for the detector with perfect efficiency. 
The expected number of events that we get agree quite well with 
the one reported in \cite{cmk}, where the results of a numerical
supernova model was used. 

If the neutrinos are massless then the time response of 
their signal at the detector
reflect just the time dependence of their luminosity function at the
source, which is the same for all the three 
flavors and hence the same for the charged current
and neutral current reactions. If neutrinos have mass $\sim eV$ then
they pick up a measurable delay during their course of flight from the 
supernova to the earth. 
For a neutrino of mass m (in eV) and energy E (in MeV), the delay
(in sec) in traveling a distance D (in 10 kpc) is
\begin{equation}
\Delta t(E) = 0.515{(m/E)}^2 D
\label{deltime}
\end{equation}
where we have neglected all the small higher order terms.
The time response curve then has contributions
from both the luminosity and the mass.  
We will now consider a scheme of neutrino masses such that 
$\Delta m_{12}^2 \sim 10^{-6} eV^2$ consistent with the solar 
neutrino problem \cite{solarmsw} and $\Delta m_{13}^2 \approx \Delta
m_{23}^2 \sim 1-10^4 eV^2$. 
The neutrino mass 
model considered here is one of several, given for the purpose 
of illustration only.  In this scheme the atmospheric neutrino anomaly 
will have to be explained by the $\numu-\nu_{sterile}$ oscillation mode 
\cite{skatm,nus}. 
The mass range for the neutrinos as the hot component of hot plus cold 
dark matter scenario in cosmology is a few $eV$ only \cite{hdm}, which 
will conflict with the higher values in the range of 
$m_{\nu_3} = 1-100 eV$ that we consider here if $\nu_3$ is stable. 
Hence, we assume that the $\nu_3$ 
state is unstable but with a large enough life time so that it is does 
not conflict with the observations of SN 1987A \cite{pbpal} 
(even though SN1987A 
observations did not correspond to any $\nutau$ event, one can put limits 
on the $\nu_3/\bar{\nu_3}$ lifetime as the $\nue/\anue$ state is a mixture 
of all the three mass eigenstates) and is 
also consistent with Big Bang Nucleosynthesis. In fact, from the 
ref. \cite{bv} we know that using the time delay technique, the 
SK and SNO can be used to probe neutrino masses down to $50 eV$ and 
$30 eV$ respectively. Hence we have presented all our results for a 
particular representative value of $m_{\nu_3} = 40 eV$. 
There have been proposals in the past for an unstable neutrino with 
mass $\sim 30 eV$ and lifetime $\sim 10^{23} sec$ \cite{sciama}. 
Since direct kinematical measurements 
give $m_\nue < 5 eV$ \cite{nue}, we have taken the $\nue$ to be 
massless and 
the charged current events experience no change. 
But since the $\nutau(\anutau)$ pick up a 
detectable time delay (for the mass spectrum of the neutrinos that we 
consider here, the $\numu(\anumu)$ do not have 
measurable time delay), the expression for the neutral current events gets 
modified to,
\begin{eqnarray}
\frac{dS_{nc}^d}{dt}&=&\frac{n}{4\pi D^2} \int dE \sigma (E)
\{N_\nue(t) f_\nue (E) + N_\anue (t)f_\anue (E) +
N_{\numu}(t)f_{\numu}(E) +
\nonumber\\
&+& N_{\anumu}(t)f_{\anumu}(E) +N_{\nutau}(t-\Delta t(E))f_{\nutau}(E) + 
N_{\anutau}(t-\Delta t(E))f_{\anutau}(E)\}
\label{del1}
\end{eqnarray}
where $dS_{nc}^d/dt$ denotes the neutral current $(nc)$ event rate with
delay $(d)$.
Delay therefore distorts the neutral current event 
rate vs. time curve. By doing a $\chi^2$
analysis of this shape distortion one can put limits on the $\nutau$ 
mass \cite{bv}. 

We next consider the neutrinos to have flavor mixing as well. 
The mixing angle $\sin^2\theta_{12}$ can be constrained from the 
solar neutrino data ($\sin^2\theta_{12} \sim 10^{-3}$) 
\cite{solarmsw} while for  
$\sin^2\theta_{13}$ there is no experimental data to fall back upon, 
but from r-process considerations in the ``hot bubble" of the 
supernova, one can restrict $\sin^2\theta_{13} \sim 10^{-6}$ \cite{qf}. 
In this scenario there will be first a matter enhanced
$\nue-\nutau$ resonance in the mantle of the supernova followed by a
$\nue-\numu$ resonance in the envelope. 
The MSW mechanism in the supernova for the neutrino mass scheme that we 
consider here is discussed in details in ref. 
\cite{qf}.
As the average energy of the $\numu/\nutau$ is greater
than the average energy of the $\nue$, neutrino flavor mixing
modifies their energy spectrum. 
Hence as pointed out in \cite{cmk}, 
the $\nue$ flux though depleted in number, gets enriched 
in high energy neutrinos and since the detection
cross-sections are strongly energy dependent, this results in the
enhancement of the charged current signal. 
The total number of events in SNO, integrated over time
in this scenario with complete flavor conversion ($P_{\nue\nue}=0$) 
are given in column 2 of Table 1. 
Of course since the $\anue$
do not have any conversion here, the $\anue$ signal remains unaltered. 
Also as the neutral current
reactions are flavor blind, the total neutral current signal remains
unchanged. But whether the time response curve of the neutral current 
signal remains unchanged in presence of mixing, in addition to delay, 
is an interesting question. 

If the neutrinos have mass as well as mixing, then the neutrinos are
produced in their flavor eigenstate, but they travel in their mass
eigenstate. The neutrino mass eigenstates
will travel with different speeds depending on their mass and will
arrive at the detector at different times.
For the scenario that
we are considering only $\nu_3$ and $\bar\nu_3$ will be delayed.
Hence to take this
delay in arrival time into account, the eq.(\ref{del1}) has to be
rewritten in terms of the mass eigenstates. 
It can be shown that expression for the neutral current event rate 
in terms of the mass eigenstates is,

\begin{eqnarray}
\frac{dS_{nc}^{do}}{dt}&=&\frac{n}{4\pi D^2} \int dE \sigma (E)
\{N_{\nu_1}(t) f_{\nu_1} (E)+N_{\bar\nu_1}(t)f_{\bar\nu_1}(E)+
N_{\nu_2}(t)f_{\nu_2}(E)
\nonumber\\
&+& N_{\bar\nu_2}(t)f_{\bar\nu_2}(E)+
N_{\nu_3}(t-\Delta t(E))f_{\nu_3}(E)+
N_{\bar\nu_3}(t-\Delta t(E))f_{\bar\nu_3}(E)\}
\end{eqnarray}
where $N_{\nu_i}$ is the $\nu_i$ flux at the source. If the neutrinos 
are produced at densities much higher than their resonance densities, 
all the mixings in matter are highly suppressed, and the neutrinos 
are produced almost entirely in their mass eigenstates. For the three 
generation case that we are considering, $\nue \approx \nu_3$, 
$\numu \approx \nu_1$ and $\nutau \approx \nu_2$. For the antineutrinos 
on the other hand, at the 
point of production in the supernova $\anue \approx \bar\nu_1$, 
$\anumu \approx \bar\nu_2$ and $\anutau \approx \bar\nu_3$.
Hence the above expression for the neutral current event rate in the 
presence of delay and mixing can be written as, 
\begin{eqnarray}
\frac{dS_{nc}^{do}}{dt}&=&\frac{n}{4\pi D^2} \int dE \sigma (E)
\{ N_\numu (t) f_\numu(E) + N_\anue(t)f_\anue(E)
+N_\nutau (t) f_\nutau(E) + N_\nutau (t) f_\nutau(E)
\nonumber\\
&+&N_\anumu (t) f_\anumu(E) + N_\anumu (t) f_\anumu(E)+
N_\nue(t-\Delta t(E))f_\nue (E)+N_{\anutau}(t-\Delta t(E))f_{\anutau}(E)\}
\label{do1}
\end{eqnarray}
Note that the above expression does not depend on the neutrino conversion 
probability as the neutral current interaction is flavor blind. 

In fig. 1 we have plotted the neutral current event rate for the reaction
($\nu_i + d \rightarrow n+p+\nu_i$, where $\nu_i$ stands for all the 6 neutrino 
species) as a function of time for massless
neutrinos along with the cases for mass but no mixing (eq.(\ref{del1}))
and mass along with mixing (eq.(\ref{do1})).
The figure looks similar for the other
neutral current reactions as well, apart from a constant normalization factor
depending on the total number of events for the process concerned. The
curves corresponding to the massive neutrinos have been given for
$m_{\nutau} = 40 eV$. 
As expected, the shape of the neutral current event rate changes due 
to the delay of massive $\nutau$. 
Since the delay given by eq.(\ref{deltime}) depends quadratically
on the neutrino mass, the distortion is more for larger masses \cite{bv2}.
But the noteworthy point is that 
the presence of mixing further distorts the rate vs. time curve. 
The reason for this distortion can be traced to the fact that 
the time delay $\propto 1/E^2$. As the energy
spectrum of the neutrinos change due to flavor mixing, 
the resultant delay is also modified and
this in turn alters the neutral current event rate as a function of time.
In fact the flavor conversion in the supernova results in 
de-energising the $\numu/\nutau$ 
spectrum and hence the delay given by eq.(\ref{deltime}) 
should increase. As larger delay caused
by larger mass results in further lowering of the neutral current event
rate vs. time curve for early times, one would normally expect that the
enhanced delay as a result of neutrino flavor conversion would have a
similar effect. But the fig. 1 shows that during the first second, the curve
corresponding to delay with mixing is higher than the one with only time
delay. This at first sight seems unexpected. But then one realizes
that while the flavor conversion reduces the average energy of the massive
$\nutau$ increasing its delay and hence
depleting its signal at early times, it energizes the 
massless and hence undelayed $\nue$ beam, which
is detected with full strength. Therefore, while for no mixing the $\nutau$
gave the larger fraction of the signal, 
for the case with mixing it is the $\nue$ that
assume the more dominant role, and so even though the $\nutau$ arrive 
more delayed compared to the case without mixing, 
the delay effect is diluted 
due to the enhancement of the $\nue$ fraction 
and the depletion of the $\nutau$ fraction of the neutral current events.  
We have also checked that although it may seem that the curve
with delay and mixing can be simulated by another curve with delay alone
but with smaller mass, the actual shape of the two curves would still be
different. This difference in shape though may not be statistically
significant and hence one may not be able to see the effect of 
mixing in the time delay of the neutrinos  
just by looking at the time response of 
the neutral current event rate 
in the present water Cerenkov detectors. We therefore look for 
various other variables which can be studied to compliment this. 

One such variable which carries information about both the neutrino mass and
their mixing is R(t), the ratio of charged to neutral current event rate as
a function of time. 
In fig. 2 we give the ratio R(t) of the total charged current to the
neutral current event rate in $\rm{D_2 O}$ in 
SNO as a function of time. Plotted are the
ratios (i) without mass, (ii) with only mixing, (iii) with delay
but zero mixing and (iv) with delay
and flavor mixing. The differences in the behavior of R(t) for the four
different cases are clearly visible. For no mass R(t)=0.3 and since the
time dependence of both the charged current and neutral current reaction
rates are the same, their ratio is constant in time. 
As the presence of mixing enhances the charged current signal keeping the 
neutral current events unaltered, R(t) goes up to 0.61 for only mixing,  
remaining constant
in time, again due to the same reason. With the introduction of delay the
ratio becomes a function of time as the neutral current reaction now has
an extra time dependence coming from the mass. At early times as the
$\nutau$ get delayed the neutral current event rate drops increasing
R(t). These delayed $\nutau$s arrive later and hence R(t) falls at
large times.
This feature can be seen for both the curves with and without mixing.
The curve for only delay starts at R(t)=0.52 at t=0 sec and falls to about
R(t)=0.26 at t=10 sec. For the delay with mixing case the corresponding
values of R(t) are 0.83 and 0.51 at t=0 and 10 sec respectively.
The important point is that the curves with and without mixing are
clearly distinguishable and should allow one to differentiate between the
two cases of only delay and delay with neutrino flavor conversion.
 
In order to
substantiate our claim that the two scenarios of only delay and delay with
mixing are distinguishable in SNO, we divide the time into bins of
size 1 second. The number of events in each bin is then used to estimate
the $\pm 1\sigma$ statistical error in the ratio R(t)
in each bin and these are then plotted in fig. 2 for the typical time bin 
numbers 1, 4 and 7. 
From the figure we see that the two cases of delay, with 
and without mixing, are
certainly statistically distinguishable in SNO for the first
6 seconds. 

We next focus our attention on $M_n^{nc}(t)$, 
the neutral current $n$-th moments of the neutrino energy 
distributions \cite{dmarc} 
observed at the detector, defined as
\begin{equation}
M_n^{nc}(t)=\int\frac{d^2S}{dEdt} E^n dE
\end{equation}
while the corresponding normalized moments are given by
\begin{equation}
\overline M_n^{nc}(t)=\frac{M_n^{nc}(t)}{M_0^{nc}(t)}
\end{equation}
We have shown the behavior of the 1st normalized moment 
$\overline M_1^{nc}(t)$ in fig. 3 as 
a function of time in SNO. For massless neutrinos, the 
$\overline M_1^{nc}$ has a 
value 40.97, constant in time, as this is again a ratio and hence 
the time dependence gets canceled out as in the case of R(t). 
For the case where the $\nutau$ is massive 
and hence delayed, it assumes a time dependence. 
Since the delay $\propto 1/E^2$ and since the neutrinos are produced 
at the source with an energy distribution, 
hence at each instant the lower 
energy $\nutau$ will be delayed more than the higher energy $\nutau$. 
Therefore $\overline M_1^{nc}(t)$, which gives the energy centroid of the 
neutral current event distribution in $\rm{D_2O}$, 
starts from a low value 38.76 at t=0 sec as all the $\nutau$ are delayed, 
rises sharply as the higher energy neutrinos arrive first 
and then falls slowly as the lower energy delayed $\nutau$ 
start arriving. If the $\nutau$ are allowed to mix with the $\nu_e$, 
then they are de-energized and the above mentioned effect is 
further enhanced. 
To make an estimate of whether SNO would be able to distinguish the 
three cases discussed above, we compute the $\pm 1 \sigma$ statistical 
errors in the $1^{st}$ normalized moment for the two scenarios 
of delay, with and without mixing, and show them for the 
$1^{st}$, $6^{th}$ and $11^{th}$ bins. We see that the errors 
involved are large enough to completely wash out the differences 
between the energy moments with and without neutrino mass and mixing. 
Hence the normalized energy moments fail to probe neutrino mass and mixing 
as at early times we don't see much difference between the different 
cases considered, while at late times the number of events become 
very small so that the error in $M_0^{nc}(t)$ becomes huge, increasing 
the error in $\overline M_1^{nc}(t)$.  

The variable that can be a useful probe 
for differentiating the case for delay 
with mixing from the case for delay without mixing is
the ratio of the unnormalized moment of the charged to neutral 
current events 
\begin{equation}
r_n(t)= \frac{M_n^{cc}(t)}{M_n^{nc}(t)}
\end{equation}
We present in fig. 4, for SNO, 
the $r_n(t)$ vs. time plots (for n=1) for the cases of 
(a) massless neutrinos (b) with mixing but no delay (c) with 
delay but no mixing and (d) with delay as well as mixing. 
Since this is a ratio, the supernova flux uncertainties get 
canceled out to a large extent and since the unnormalized moments 
have smaller statistical errors, this is a better variable than the 
normalized moments to observe the signatures of neutrino mixing. In the 
figure we have shown the $\pm 1 \sigma$ statistical errors in $r_1(t)$ 
for the two cases of delay alone and delay with mixing, for the 
$1^{st}$, $8^{th}$ and $15^{th}$ bins in time, and the two cases are 
clearly distinguishable in SNO for early as well as late times. 
Note that $r_1(t)$ is different from the ratio R(t) as it gives 
information about the ratio of the energy centroids of the 
charged current and neutral current distributions as a function 
of time, while the latter gives only the ratio of the number of events 
as a function of time.  

The advantage of using ratios 
is that, they are not only sensitive to the mass and
mixing parameters but are also almost insensitive to the details of 
supernova models. 
Since they are a ratio they are almost
independent of the luminosity and depend only
on some function of the ratio of neutrino temperatures. 
All the calculations presented so far are for fixed neutrino 
temperatures. In 
order to show that the time dependence of the neutrino temperatures 
does not alter our conclusions much, we present our analysis
with time dependent neutrino temperatures. We take
\begin{equation}
T_\nue = 0.16\log t + 3.58,~~
T_\anue= 1.63 \log t + 5.15,~~
T_\numu= 2.24 \log t + 6.93
\end{equation}
These forms for the neutrino temperatures follow from fits to the results of
the numerical supernova model given in Totani {\it et al.} \cite{totani}. In
fig. 5 we compare the ratio R(t) for the cases of delay and
delay with mixing for the two cases
of fixed temperatures and the time dependent temperatures. 
It is clear from the figure that
that the time dependence of the neutrino temperatures does not have
much effect on the time dependence of the ratio of the
charged current to neutral current rates. In fact the two curves
corresponding to fixed and time dependent temperatures, fall within
$\pm 1\sigma$ statistical errorbars for both the cases of only delay 
and delay with mixing.

In conclusion, 
we have shown that even though neutrino flavor mixing cannot alter the 
total neutral current signal in the detector - the neutral current 
interaction being flavor blind, it can have a non-trivial impact on 
the delay of massive neutrinos, which alters the neutral current event 
rate as a function of time. The neutral current event rate though 
does not depend on the neutrino conversion probability. 
In order to study the effect of neutrino mass and mixing we have suggested  
various variables. Of the different variables that we have presented here, 
the ratio of the charged to neutral current event rate R(t), can show 
the effect of mixing during the first few seconds, while the 
charged to neutral 
current ratio of the energy moments 
are useful diagnostic tools for all times. 
These variables are not just sensitive 
to flavor mixing and time delay, they are also insensitive to 
supernova model uncertainties and hence are excellent tools to study 
the effect of flavor mixing on the time delay of massive supernova 
neutrinos.  

In this letter we have considered a mass spectrum for the neutrinos 
where only the $\nutau$ have a measurable delay. 
The model considered is one of many,  
but one can easily 
extend the above formalism to include 
more general classes of neutrino models \cite{ck}. 
In addition to the energy moments that we have presented here, the 
$n$-th order moments of the arrival time of the neutrinos 
as a function of energy can also be 
analyzed to study the effect of neutrino mass and mixing, and we plan 
to present them in a future work \cite{ck}. 

\vskip 4mm
{\it The authors thank S. Goswami, P.B. Pal, S. Mohanty 
and D. Majumdar for discussions.}

\newpage

\begin{center}
{\bf Table 1}
\end{center}
The expected number of neutrino events in SNO. To get
the number of events in SK, one has to scale the number of events in
$\rm{H_2O}$ given here to its fiducial mass of 32 kton.
The column A corresponds to massless neutrinos,
column B to neutrinos with complete flavor conversion
The $\nu_i$ here refers to all the six neutrino species.

\[
\begin{tabular}{|c| c c |} \hline
{reactions in 1 kton $\rm{D_2O}$} & {\hspace{5mm} A\hspace{5mm}} &
{\hspace{5mm} B\hspace{5mm} } \\
\hline

{$\nu_e+d\rightarrow p+p+e^-$} & { 75 } & {239} \\ \hline
{$\bar\nu_e +d\rightarrow n+n+e^+$} & {91} & {91} \\ \hline
{$\nu_i+d\rightarrow n+p+\nu_i$} & {544} & {544} \\ \hline
{$\nu_e +e^- \rightarrow \nu_e +e^-$} & {4} & {6} \\ \hline
{$\bar\nu_e+e^- \rightarrow \bar\nu_e+e^-$}
& {1} & {1} \\ \hline
{$\nu_{\mu,\tau}(\bar\nu_{\mu,\tau}) +e^- \rightarrow
\nu_{\mu,\tau}(\bar\nu_{\mu,\tau}) +e^-$}
& {4} & {3} \\ \hline
{$\nu_e +^{16}O \rightarrow e^- +^{16}F$} & {1} & {55} \\ \hline
{$\bar\nu_e + ^{16}O\rightarrow e^+ + ^{16}N$} & {4} & {4} \\ \hline
{$\nu_i +^{16}O \rightarrow \nu_i +\gamma +X$} & {21} & {21} \\ \hline

{reactions in 1.4 kton $\rm{H_2O}$} & {} & {} \\
\hline

{$\bar\nu_e +p\rightarrow n+e^+$} & {357} & {357} \\ \hline
{$\nu_e +e^- \rightarrow \nu_e +e^-$} & {6} & {9} \\ \hline
{$\bar\nu_e+e^- \rightarrow \bar\nu_e+e^-$} & {2} & {2} \\ \hline
{$\nu_{\mu,\tau}(\bar\nu_{\mu,\tau}) +e^- \rightarrow
\nu_{\mu,\tau}(\bar\nu_{\mu,\tau}) +e^-$} & {6} & {5} \\ \hline
{$\nu_e +^{16}O \rightarrow e^- +^{16}F$} & {2} & {86} \\ \hline
{$\bar\nu_e + ^{16}O\rightarrow e^+ + ^{16}N$} & {6} & {6} \\ \hline
{$\nu_i +^{16}O \rightarrow \nu_i +\gamma +X$} & {33} & {33} \\ \hline
\end{tabular}
\]


\begin{center}
{\bf Figure Captions}
\end{center}

\noindent
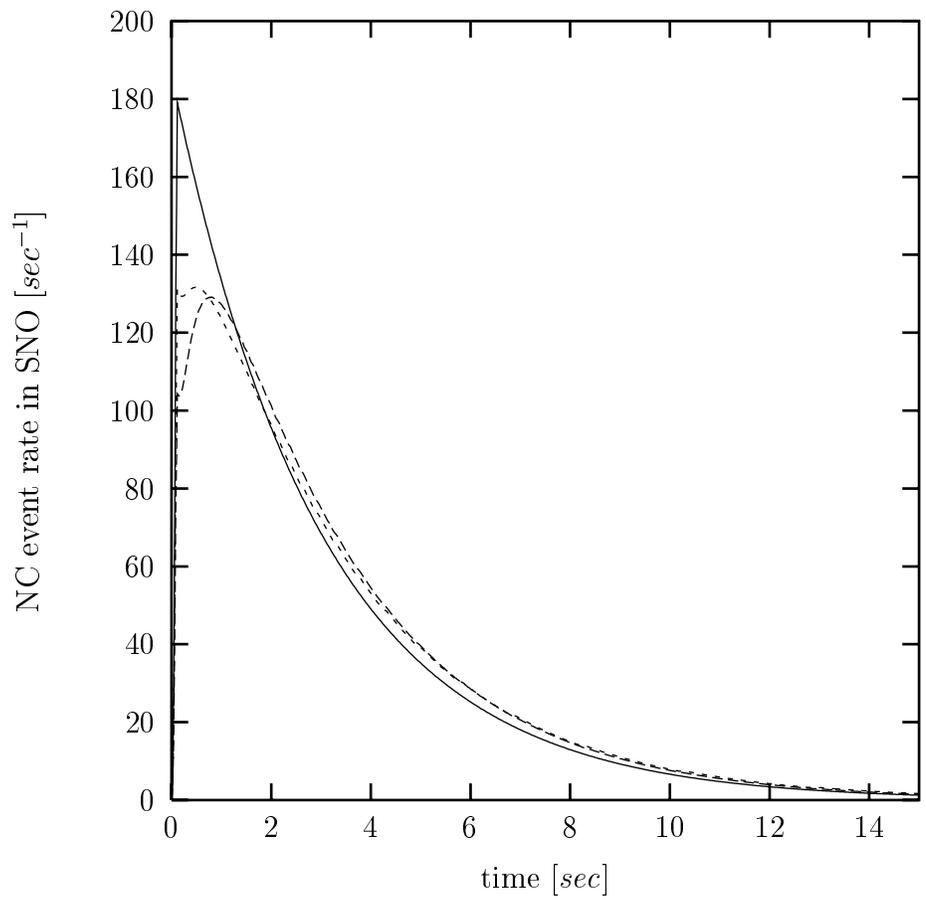
{\bf Fig.1} The neutral current event rate as a function of time in $\rm D_2O$ 
in SNO. The solid line corresponds to the case of
massless neutrinos, the long dashed line to neutrinos with only mass but
no mixing, while the short dashed line gives the event rate for neutrinos
with mass as well as flavor mixing. 

\vskip 0.5 cm

\noindent
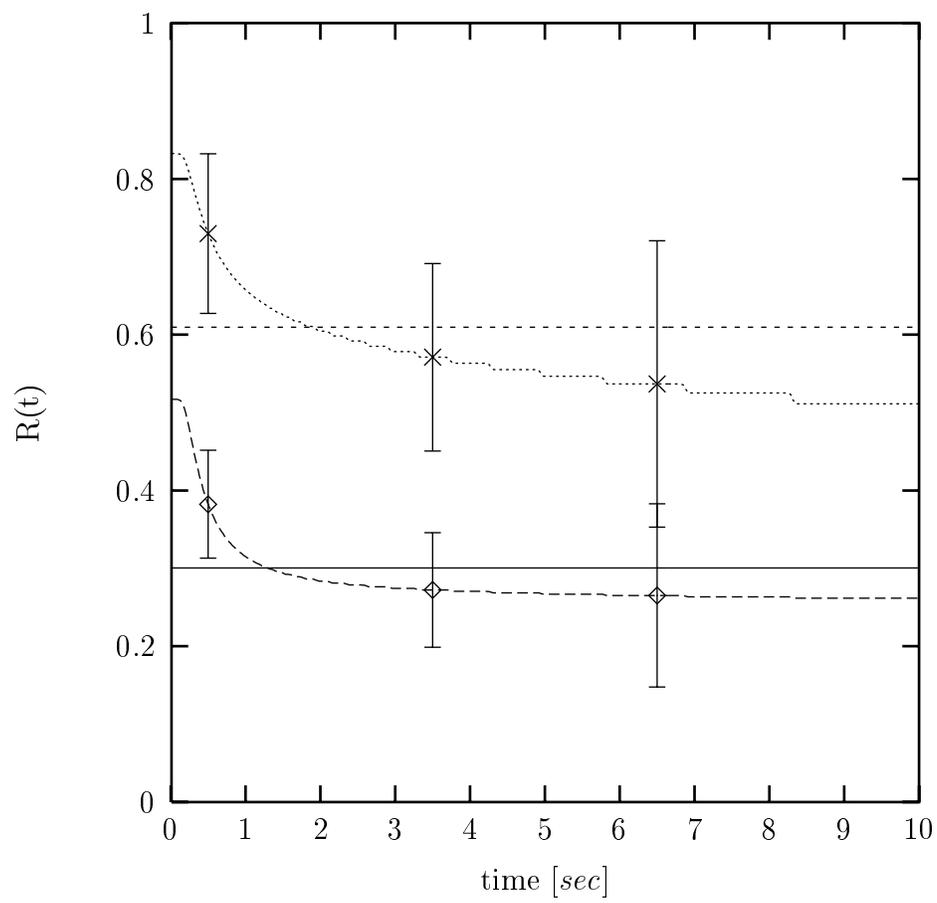
{\bf Fig.2} The ratio R(t) of the total charged current to neutral current
event rate in SNO versus time. The solid line is for massless
neutrinos, the short dashed line for neutrinos with complete 
flavor conversion but
no delay, the long dashed line for neutrinos with only delay and no flavor
conversion and the dotted line is for neutrinos with both delay and 
complete flavor conversion.
Also shown are the $\pm 1\sigma$ statistical errors for 
delay with and without mixing in the $1^{st}$, $4^{th}$ and the 
$7^{th}$ time bins. 

\vskip 0.5 cm
\noindent
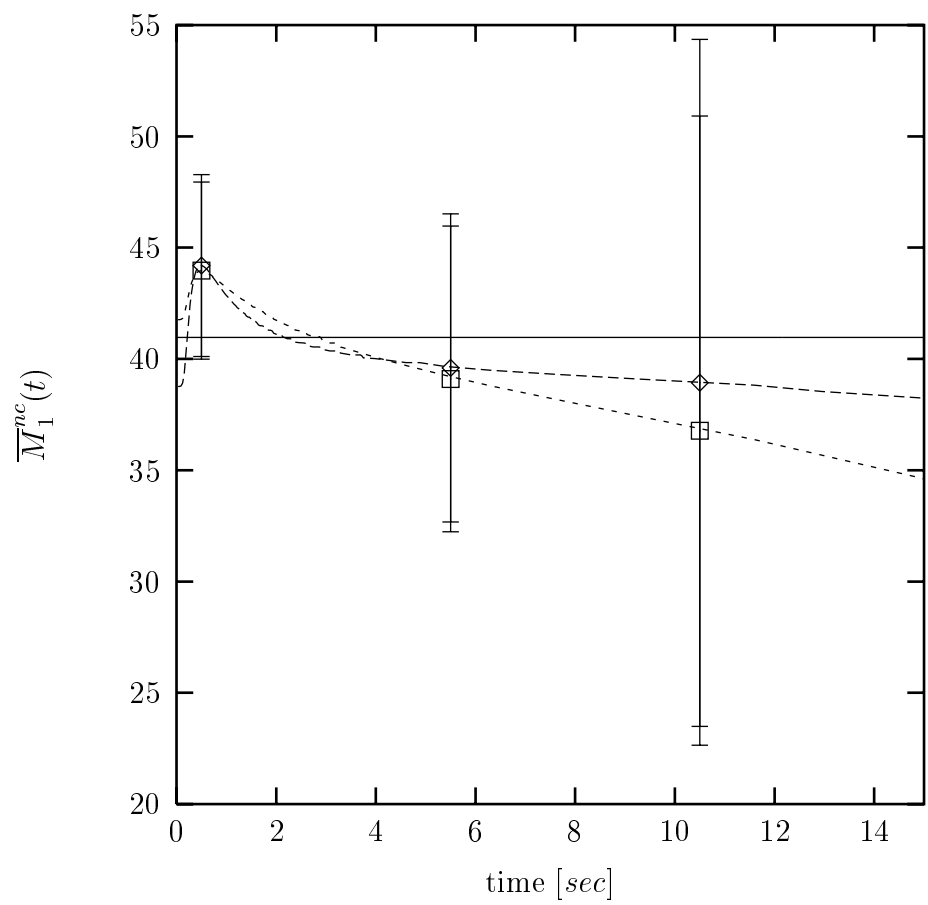
{\bf Fig. 3} The 1st normalized energy moment of the neutral current 
events in SNO $\overline M_1^{nc}(t)$ versus time. The solid line 
corresponds to the case of
massless neutrinos, the long dashed line to neutrinos with only mass but
no mixing, while the short dashed line gives the event rate for neutrinos
with mass as well as flavor conversion.
Also shown are the $\pm 1\sigma$ statistical errors for 
delay with and without mixing in the $1^{st}$, $6^{th}$ and the 
$11^{th}$ time bins. 

\vskip 0.5 cm

\noindent
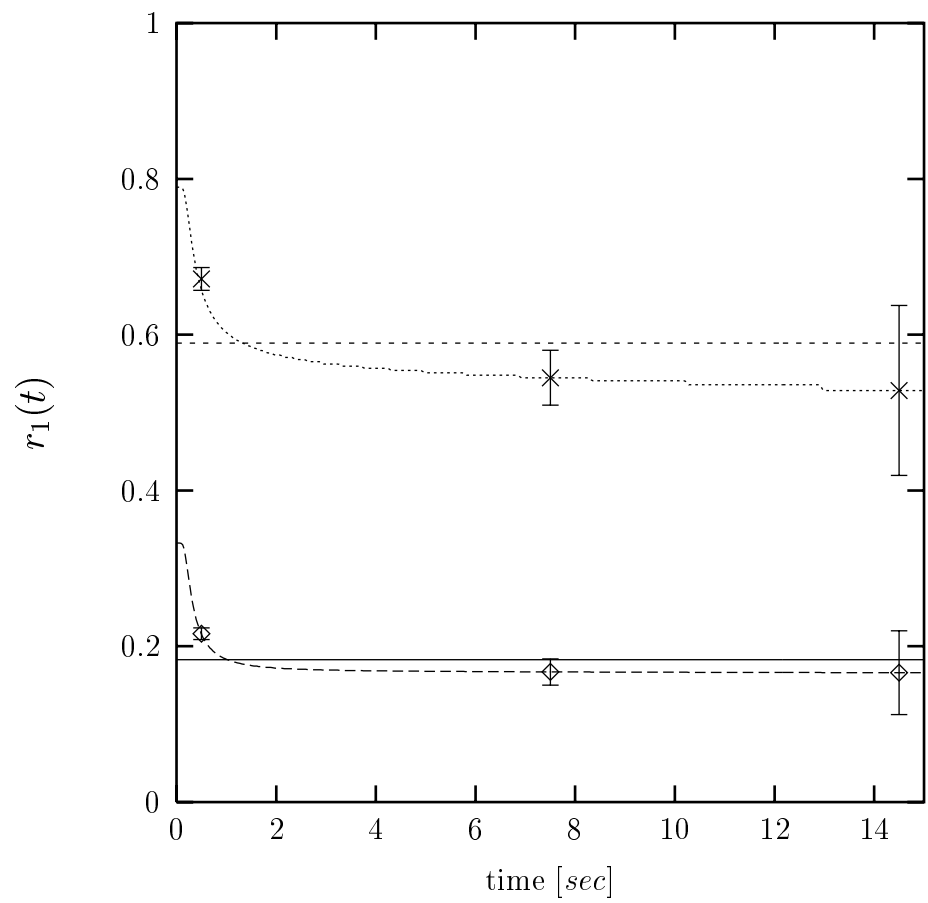
{\bf Fig. 4} The variation of $r_1(t)$ with time in SNO. 
The solid line is for massless
neutrinos, the short dashed line for neutrinos with complete
flavor conversion but
no delay, the long dashed line for neutrinos with only delay and no flavor
conversion and the dotted line is for neutrinos with both delay and  
complete flavor conversion.
Also shown are the $\pm 1\sigma$ statistical errors for 
delay with and without mixing in the $1^{st}$, $8^{th}$ and the 
$15^{th}$ time bins. 

\vskip 0.5 cm

\noindent
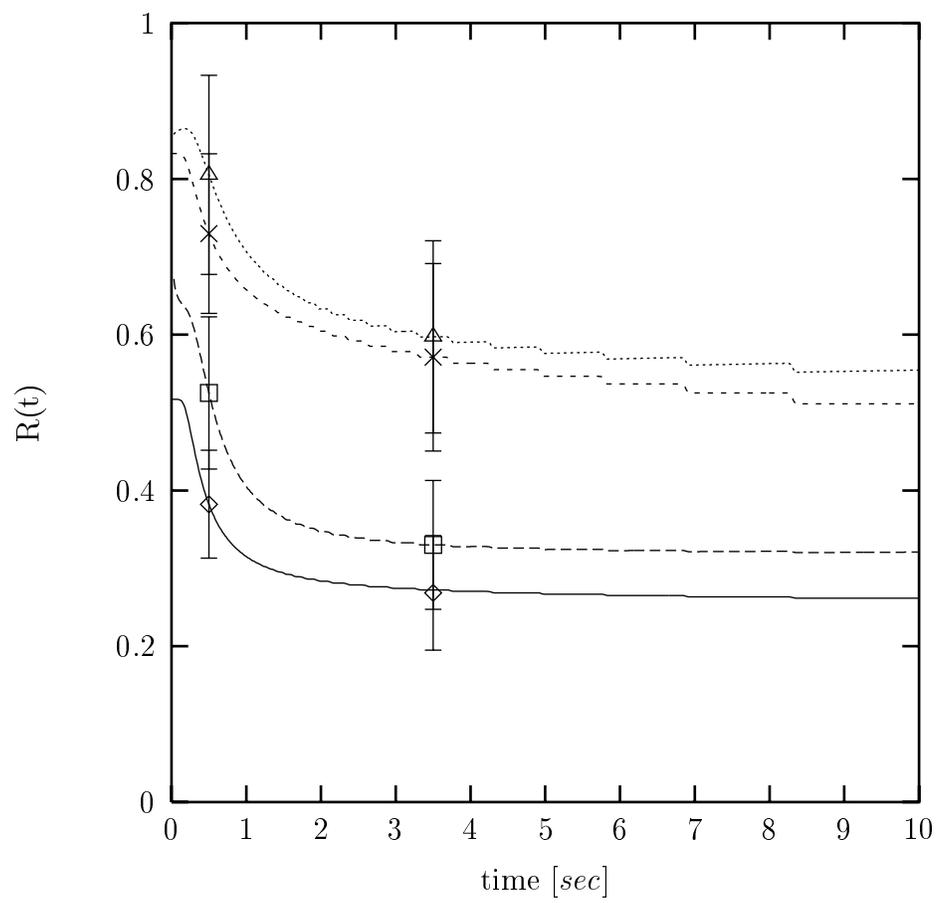
{\bf Fig. 5} The ratio R(t) in SNO for the two cases of fixed
and time dependent neutrino temperatures. The solid line and the long
dashed line give R(t) for the cases of fixed temperatures
and varying temperatures respectively for only delay,
while the short dashed line and the dotted line
give the corresponding R(t) for delay with mixing. 
We have also given the $\pm 1\sigma$ statistical errors 
in the $1^{st}$ and the $4^{th}$ time bin, 
for the both the curves for fixed and time dependent temperatures.

\end{document}